\documentclass[12pt]{iopart}
\usepackage{iopams}
\usepackage{amssymb}
\usepackage{bm}
\def\Quadrat#1#2{{\vcenter{\hrule height #2
\hbox{\vrule width #2 height #1 \kern#1
\vrule width #2}\hrule height #2}}}
\def\Box{\mathop{\kern 1pt\hbox{$\Quadrat{8pt}{0.4pt}$} \kern 1pt}}
\begin{document}
\comment[Comment on the "speed
of gravity/speed of light" controversy]{Comment on 'Model-dependence of Shapiro time delay and the "speed
of gravity/speed of light" controversy'}
\author{Sergei M. Kopeikin}
\address{Department of Physics \& Astronomy, University of
Missouri-Columbia, Columbia, MO 65211, USA}
\ead{kopeikins@missouri.edu}
\begin{abstract}
In a recent paper published in {\it Class. Quant. Grav.}, 2004, {\bf 21}, 3803  Carlip used a vector-tensor theory of gravity to calculate the Shapiro time delay by a moving gravitational lens. He claimed that the relativistic correction of the order of $v/c$ beyond the static part of the Shapiro delay depends on the speed of light $c$ and, hence, the Fomalont-Kopeikin experiment is not sensitive to the speed of gravity $c_g$. In this letter we analyze Carlip's calculation and demonstrate that it implies a gravitodynamic (non-metric) system of units based on the principle of the constancy of the speed of gravity but it is disconnected from the practical method of measurement of astronomical distances based on the principle of the constancy of the speed of light and the SI metric (electrodynamic) system of units. Re-adjustment of theoretically-admissible but practically unmeasurable Carlip's coordinates to the SI metric system of units used in JPL ephemeris, reveals that the velocity-dependent correction to the static part of the Shapiro time delay does depend on the speed of gravity $c_g$ as shown by Kopeikin in {\it Class. Quant. Grav.}, 2004, {\bf 21}, 1. This analysis elucidates the importance of employing the metric system of units for physically meaningful interpretation of gravitational experiments.
\end{abstract}
\submitto{\CQG}
\maketitle

The solar system can serve as a laboratory testbed of general relativity if and only if the distances and velocities of the solar system bodies are known in the SI metric system of units \cite{nist}. Their precise determination is the main goal of the fundamental (ephemeris) astronomy that presently uses optical measurements of the Sun and planets in combination with ultra-precise radar and laser ranging to spacecrafts and retro-reflectors on the Moon \cite{standish,pitjeva}. It is critical to realize that astronomers conducting gravitational experiments can not use rigid sticks and rulers for measuring distances between the solar system bodies. The only practical way to do it, is to use electromagnetic signals (light). According to general relativity light moves in the curved space-time along null geodesics, the observation of which can be used to study various physical properties of the gravitational field. The main principle of measuring astronomical distances is the constancy and isotropy of the speed of light in the Minkowskian space which (according to general relativity) is a tangent space attached to each point of the curved space-time manifold \cite{LL,mtw,brum}. This principle was confirmed in a multidude of ingenious laboratory experiments \cite{zhang} and is so fundamental that since 1983 the distances are practically measured in terms of the SI meter, which is defined as the length of the path travelled by light in vacuum during a time interval of 1/299 792 458 of a second \cite{nist}. This definition means that the speed of light in the SI metric system of units is exactly $299 792 458$ m$\cdot$s$^{-1}$. We shall work in units with the speed of light equal to unity, so that distances are measured in light seconds. In this system of units the definition of SI meter is as follows 
\begin{equation}\label{simeter}
[1\;{\rm m}\;]_{_{\rm SI}}= \left[\frac{1\;{\rm s}}{c}\right]\;,
\end{equation} 
where $1\;{\rm s}$ is one SI second, and $c=299 792 458$ denotes the (dimensionless) numerical value of the speed of light \cite{nist}.

General relativity is a theory of gravity that is conceptually independent of Maxwell's theory of electrodynamics. Therefore, vacuum Einstein's equations, $G_{\mu\nu}=0$, could describe a free gravitational field propagating with the fundamental speed $c_g$ being different from the speed of propagation of light $c$ that obeys the Maxwell equations. In other words, the tangential space to the Einsteinian space-time manifold with the gravitational field residing on it could be different from the Minkowskian space-time of the Maxwellian electrodynamics. Being conceptually independent from Maxwell's theory the general theory of relativity (and any other self-consistent theory of gravity) would admit a practical realization of a gravitodynamic (GD) unit of length (gravitodynamic meter) based on the constancy of the speed of propagation of gravitational waves, $c_g$. The gravitodynamic and SI meter could differ as they are built up out of different physical principles. However, general relativistic paradigm approves $c_g=c$ and, hence, does not distinguish the two definitions of meter. This paradigm can be violated in some of alternative theories of gravity and its testing is to be one of the main goals of experimental gravitational physics \cite{w-book}. 

One way to achieve this goal is to study propagation of light in curved space-time which has a dynamic character, for example, a moving gravitational lens (Jupiter) \cite{k-apj}. To analyze this problem Carlip \cite{carlip} uses a vector-tensor theory of gravity in which the time-dependent components of the gravitational field (the metric tensor and the Christoffel symbols) of the lens are normalized to the speed of gravity $c_g$ that is a universal constant characterizing both the speed of propagation of weak gravitational waves and gravitomagnetic phenomena including the Lorentz invariance of the metric tensor perturbations \cite{k-cqg,k-grqc}. The strength of the gravitomagnetic coupling between the rotational spin of the Earth and that of a free-falling gyroscope is currently under experimental verification by the GP-B space mission \cite{gpb}. The Lorentz invariant property (= causality) of the gravitational field due to its finite speed of propagation has been recently corroborated in VLBI experiment conducted by Fomalont and Kopeikin \cite{fk-apj} who confirmed the equality  $c_g=c$ with the precision 20\%. The "speed of gravity" interpretation of the experimental results of \cite{fk-apj} is based on the original publication by Kopeikin \cite{k-apj} (see also \cite{k-cqg,k-grqc,k-pla,k-ijmp}). This interpretation was antagonized by some researchers arguing that the Fomalont-Kopeikin experiment measured the speed of light incoming to observer from the quasar \cite{willsite}. Carlip's paper \cite{carlip} represents the most advanced attempt in support of the quasar "speed of light" interpretation and we basically focus on the discussion of Carlip's arguments.

Carlip \cite{carlip} calculated the gravitational time delay, $\Delta$, for light propagating through the field of a uniformly moving massive body (gravitational lens) in the framework  of a vector-tensor theory of gravity that leads to the existence of two metric tensors, $g_{\alpha\beta}$ and $\tilde{g}_{\alpha\beta}$, describing propagation of gravity and light with the speeds of $c_g$ and $c$ respectively. Carlip obtained 
\begin{equation}\label{1}
\Delta = - \frac{(1+\gamma)Gm}{c^3}\left\{
  \left[ 1 - (1+\zeta)\frac{{\bm k}\cdot{\bm v}_J}{c}\right]\ln
  \left(r - {\bm K}\cdot{\bm r}\right)\right\}\;,
\end{equation}
where $c$ is the speed of light, $\gamma$ and $\zeta$ are parameters of the parametrized post-Newtonian (PPN) formalism \cite{w-book}, the unit vector  
\begin{equation}\label{2}
{\bm K} = {\bm k} - \frac{1}{c}{\bm k}\times({\bm v}_J\times{\bm k})\;, 
\end{equation}
vector ${\bm r}={\bm x}-{\bm x}_J$ describes the difference between the spatial coordinates ${\bm x}$ of observer and the moving light-ray deflecting body (Jupiter), ${\bm x}_J$, both taken at the time of observation $t$, $r=|{\bm r}|$, velocity of Jupiter ${\bm v}_J=d{\bm x}_J/dt$, and ${\bm k}$ is the unit vector in the direction of the incoming light ray from a radio source (quasar). It is urgent to emphasize that Carlip \cite{carlip} works with space-time coordinates $x^\alpha=(x^0,{\bm x})$, where $x^0=c_gt$, and the speed of gravity $c_g$ is considered as a primary fundamental constant. This assumption implicitly suggests that 
coordinates ${\bm x}_J$ and velocity ${\bm v}_J$ of Jupiter (and other planets and Sun) are expressed in a non-metric (gravitodynamic) system of units where the speed of gravity $c_g$ is assumed to be known while the speed of light $c$ is an unknown parameter that is to be determined experimentally. In his paper \cite{carlip} Carlip assumes the speed of gravity equal to unity (see remark given immediately after equation (2.2) in \cite{carlip}). Because of this assumption the speed of gravity appears nowhere explicitly in the rest of the paper \cite{carlip} as it is absorbed into the definition of the gravitodynamic unit of length -- the gravitodynamic (GD) meter -- which is based on the speed of propagation of weak gravitational waves 
\begin{equation}\label{carlipmeter}
[1\;{\rm m}\;]_{_{\rm GD}}= \left[\frac{1\;{\rm s}}{c_g}\right]\;.
\end{equation} 
The gravitodynamic meter is conceptually different from the (electrodynamic) SI meter  (\ref{simeter}) and it makes interpretation of gravitational measurements in the vector-tensor theory of gravity more convoluted.

Relationship between the gravitodynamic and SI meters is derived from comparison of equations (\ref{simeter}) and (\ref{carlipmeter})
\begin{equation}\label{cgc}
[1\;{\rm m}\;]_{_{\rm GD}}= \frac{c}{c_g}\,[1\;{\rm m}\;]_{_{\rm SI}}\;.\end{equation} 
The main goal of the Fomalont-Kopeikin experiment \cite{fk-apj} was to measure the ratio $\epsilon=c/c_g$. We assumed that the numerical value of the speed of light, $c$, is known so that the experiment measured $c_g$. Carlip \cite{carlip} assumed that the speed of gravity, $c_g$ is known and the experiment measured $c$. We emphasize that Carlip's
assumption is plausible but incompatible with real practice since it assumes that the speed of gravity $c_g$ can be somehow measured alone, independently of the speed of light $c$, and used for establishing a working standard of the gravitodynamic meter (\ref{carlipmeter}). But this is impossible because the gravitational waves have not been discovered as yet and the electromagnetic field is the only fundamental field that can be used for direct technological measurement of distances to and velocities of astronomical bodies in the solar system.

Kopeikin \cite{k-cqg,k-apj,k-pla} introduced the speed of gravity parametrization of the Einstein equations of general relativity to distinguish relativistic effects associated with the speed of gravity $c_g$ from those normalized to the speed of light $c$ which is assumed to be known in accordance with the real practice of astronomical measurements. He has derived for the $c_g$-parametrized gravitational time delay the following equation \cite{k-cqg,k-pla}
\begin{equation}\label{3}
\Delta = - \frac{2Gm}{c^3}\left\{
  \left[ 1 - \frac{{\bm k}\cdot{\bm V}_J}{c_g}\right]\ln
  \left(R - {\bm K}\cdot{\bm R}\right)\right\}\;,
\end{equation}
where the unit vector  
\begin{equation}\label{4}
{\bm K} = {\bm k} - \frac{1}{c_g}{\bm k}\times({\bm V}_J\times{\bm k})\;, 
\end{equation}
radius-vector ${\bm R}={\bm X}-{\bm X}_J$ describes the difference between the spatial coordinates of observer,  ${\bm X}$, and Jupiter, ${\bm X}_J$, both taken at the time of observation $t$, $R=|{\bm R}|$, velocity of Jupiter ${\bm V}_J=d{\bm X}_J/dt$,  and one takes the PPN parameters $\gamma=1$, $\zeta=0$ as in general relativity since their numerical values are not essential for the following discussion. In our approach \cite{k-cqg,k-pla} the coordinates, ${\bm X}_J$, and velocity, ${\bm V}_J$, of Jupiter are expressed in the SI metric system of units where the speed of light is known. In the present paper we have used capitalized letters for these coordinates and velocities in order to distinguish them from Carlip's coordinates ${\bm x}_J$ and velocities ${\bm v}_J$ which are normalized to the gravitodynamic meter (\ref{carlipmeter}). 

The speed of light $c$ in equation (\ref{2}) is unknown and considered as a measurable parameter in Carlip's approach \cite{carlip}. Its measurement requires to know precisely the velocity of Jupiter ${\bm v}_J$. However, this velocity is given in \cite{carlip} in the non-metric (gravitodynamic) unit of length. This unit is convenient for theoretical calculations in the vector-tensor theory of gravity \cite{carlip} but it is not practically available in astronomical measurements because astronomers use light to measure coordinates and velocities of the solar system bodies \cite{standish,pitjeva,brum}. Hence, we can not directly implement Carlip's formulation (\ref{1}), (\ref{2}) of the time delay to process the data of either the Fomalont-Kopeikin or any other gravitational experiments. What is available in reality through the JPL ephemeris \cite{standish} are coordinates ${\bm X}_J$ and velocity ${\bm V}_J$ of Jupiter, measured in the SI metric system of units which are known from laboratory measurements. Relationship between the two systems of the coordinates and velocities is established by equation (\ref{cgc})
\begin{equation}\label{5}  
{\bm x}_J=\frac{c}{c_g}{\bm X}_J\;,\qquad\qquad {\bm v}_J=\frac{c}{c_g}{\bm V}_J\;,
\end{equation}
which yields, ${\bm v}_J/c={\bm V}_J/c_g$, so that the equations (\ref{2}) and (\ref{4}) are, in fact, identical. It makes clear that Carlip's equation (\ref{2}) describes the aberration of light in the gravitodynamic system of units while Kopeikin's equation (\ref{4}) describes the aberration of gravity in the metric system of units. Carlip \cite{carlip} holds the speed of gravity $c_g$ hidden to the definition of the velocity of Jupiter ${\bm v}_J$ which is measured in the gravitodynamic meters (\ref{carlipmeter}) divided by SI second but these meters are not used in JPL ephemeris \cite{standish} so that ${\bm v}_J$ remains unknown and practically useless. 

Carlip \cite{carlip} (see also \cite{carlip2,w-apj}) discusses a slow motion (post-Newtonian) expansion of the Lienard-Wiechert solution of the wave equation for a gravitational potential $\phi$ generated by a point mass (Jupiter) moving along a world line ${\bm x}_J(t)$ 
\begin{equation}\label{6}
\left[-\frac{1}{c^2_g}\frac{\partial^2}{\partial t^2}+\nabla^2\right]\phi(t,{\bm x})=-4\pi G M_J\delta\left({\bm x}-{\bm x}_J(t)\right)\;,
\end{equation}
where $G$ is the universal gravitational constant, $\nabla^2$ is the Laplace differential operator, $M_J$ is mass of Jupiter, and $\delta({\bm x})$ is Dirac's delta-function. Retarded (causal) solution of equation (\ref{6}) is
\begin{equation}\label{7}
\phi(t,{\bm x})=\frac{GM_J}{\rho-c_g^{-1}{\bm v}_J\cdot{\bm\rho}}\;,
\end{equation}
where ${\bm\rho}={\bm x}-{\bm x}_J(s)$, $\rho=|{\bm\rho}|$, and position of Jupiter is taken at the retarded instant of time 
\begin{equation}\label{8}
s=t-\frac{1}{c_g}\rho\;.
\end{equation}
The post-Newtonian expansion of the right side of equation (\ref{7}) with respect to the ratio $v_J/c_g$ yields
\begin{equation}\label{9}
\phi(t,{\bm x})=\frac{GM_J}{|{\bm x}-{\bm x}_J(t)|}+O\left(\frac{v_J^2}{c_g^2}\right)\;,
\end{equation}
that shows that the speed of gravity $c_g$ appears explicitly only in the quadratic residual terms. This is the main physical argument of Carlip \cite{carlip} against the "speed of gravity" interpretation of the results of the Fomalont-Kopeikin experiment which measured terms of the linear order \cite{fk-apj}. 

However, this argument has a hidden pitfall which has been missed by Carlip \cite{carlip}. Fact of the matter is that the coordinates of Jupiter, ${\bm x}_J(t)$, used by Carlip \cite{carlip} in equation (\ref{9}) depend implicitly on the speed of gravity $c_g$ since they are expressed in the gravitodynamic unit of length (\ref{carlipmeter}) utilizing propagation of weak gravitational waves. In order to calculate the Shapiro time delay in the field of moving Jupiter one makes use of a linear expansion 
\begin{eqnarray}\label{linex}
{\bm x}_J(t)={\bm x}_J(t_A)+{\bm v}_J(t-t_A)
=\frac{c}{c_g}\left[{\bm X}_J(t_A)+{\bm V}_J(t-t_A)\right]\;,
\end{eqnarray}
where $t_A$ is a fiducial instant of time, ${\bm x}_J(t_A)$ is the initial (constant) position of Jupiter, and we used equation (\ref{5}) to convert unobservable coordinates ${\bm x}_J$ and velocity of Jupiter ${\bm v}_J$ to their JPL ephemeris-based counterparts: ${\bm X}_J$ and ${\bm V}_J$.    
The unperturbed light-ray trajectory 
\begin{equation}\label{kok}
{\bm x}=c{\bm k}(t-t_0)+{\bm x}_0\;,
\end{equation} 
depends only on the speed of light $c$, so that after substitution into equation (\ref{9}) 
along with expansion (\ref{linex}) one can compare (measure) the speed of gravity $c_g$ versus speed of light $c$ which is used as a reference standard known from laboratory measurements. Notice that the speed of gravity $c_g$ drops out from the Newtonian equations of motion of celestial bodies because in this case coordinate ${\bm x}$ in (\ref{9}) is a coordinate of another massive body which has the same Taylor expansion as shown in equation (\ref{linex}). Thus, the ratio $c/c_g$ gets factorized and simply re-defines the universal gravitational constant, thus, becoming unmeasurable in the Newtonian celestial mechanics of astronomical bodies. Factorization of $c/c_g$ also takes place in the static part of the Shapiro delay making $c_g$ unmeasurable in the reference frames where the gravitational field is represented as time-independent.

 The integration of the gravitationally perturbed light geodesic equation with equations (\ref{linex}) and (\ref{kok}) taken into account, leads then \cite{k-cqg,k-pla} to equation (\ref{4}). The conclusion is that the Newtonian part of the gravitational potential $\phi$ implicitly contains information about the speed of gravity encoded in the velocity of the massive body in terms of the (gravitodynamic) units of its measurement comparatively to the SI units. This is because Carlip's vector-tensor theory of gravity uses the Einstein procedure for synchronization of events in curved space-time manifold based not on the electromagnetic signals but on the propagation of weak gravitational waves. Self-consistency of such gravitational-wave synchronization with that based on the exchange of electromagnetic signals can be checked by measuring high-order relativistic effects in propagation of light in time-dependent gravitational fields which allows us to compare the gravitodynamic versus metric system of units, that is the ratio $\epsilon=c/c_g$ \cite{fk-apj}. 

We finally emphasize that propagation of light through the field of moving Jupiter tests the dynamic general-relativistic effect of the next order of magnitude beyond the first post-Newtonian (1 PN) approximation, which is not conceptually reduced to the ordinary measurement of the aberration of light in flat space-time \cite{w-apj} or the Roemer delay as mistakenly written in \cite{paskual}. Indeed, the Lorentz covariant general-relativistic time delay equation reads \cite{ks,kf-abb}
\begin{equation}
\label{10}
t-t_0=\frac{1}{c}{\bm k}\cdot\left({\bm x}-{\bm
x}_0\right)-\frac{2GM_J}{c^3}\frac{1-c^{-1}{\bm k}\cdot{\bm
v}_J}{\sqrt{1-v^2_J/c^2}}\ln\left(\rho-{\bm k}\cdot{\bm\rho}\right)\;,
\end{equation}
where ${\bm\rho}={\bm x}-{\bm x}_J(s)$, $\rho=|{\bm \rho}|$, and the retarded
time $s$ is defined by the gravity cone equation (\ref{8}) with $c_g=c$ in accordance with general relativity \cite{LL,mtw}. 

Appearance of the retarded position ${\bm x}_J(s)$ of Jupiter in equation (\ref{10}) has a simple physical explanation. The retarded solution (\ref{7}) of the wave equation (\ref{6}) describes two effects -- the retardation and the aberration of gravity \cite{carlip2}. Because they are equal in the linear order of the post-Newtonian expansion these effects cancel each other out in this approximation. However, when one uses the retarded solution  (\ref{7}) of the wave equation (\ref{6}) for calculating time delay of light caused by time-dependent gravitational field one more physical effect appears -- the aberration of light. In general relativity the aberration of light and the aberration of gravity are also equal to each other in the linearized approximation and, hence, compensate each other in the final expression (\ref{10}), thus, leaving the retardation of gravity effect observable in the form of Jupiter's retarded position ${\bm x}_J(s)$. All this means that the Fomalont-Kopeikin experiment \cite{fk-apj} is a null-type experiment testing the local equivalence between the null cones of the gravitational and electromagnetic waves which are defined by the speed of gravity, $c_g$, and that of light, $c$, respectively. 

\ack
We are thankful to G.E. Melki for careful reading of the manuscript and valuable comments. This work has been supported by the Eppley Foundation for Research (New York) and the Research Council of the University of Missouri-Columbia.

\end{document}